\documentclass[11pt,a4paper]{article}
\usepackage[utf8]{inputenc}
\usepackage{graphicx}
\usepackage{subcaption}
\usepackage{amsmath,amssymb,mathtools}
\usepackage{hyperref}
\hypersetup{colorlinks=true,linkcolor=blue,citecolor=blue,urlcolor=blue}
\usepackage{geometry}
\date{}
\begin{document}
\title{{\bf Landauer erasure and horizon information in an expanding universe}}

\author{Babak Vakili\thanks{email: ba.vakili@iau.ac.ir}\\\\{\small {\it Department of Physics, CT.C., Islamic Azad
University, Tehran, Iran}}} \maketitle

\begin{abstract}
We investigate the thermodynamic cost of information erasure in
Friedmann--Robertson--Walker cosmology by combining Landauer's
principle with the thermodynamics of the apparent horizon. Adopting
the apparent-horizon temperature as an effective cosmological
temperature, we find that for standard non-phantom expansion, the minimum
heat required to erase one bit decreases while the entropy-equivalent
information content of the horizon increases. We further compare the Landauer heat with the
Misner--Sharp energy and show that, despite their simple thermodynamic
relation, they have distinct physical meanings. These results provide
a simple connection between information erasure, horizon
thermodynamics, and gravitational energy in an expanding universe.
\vspace{5mm}\noindent
\end{abstract}

\section{Introduction}

Information has evolved from a mathematical concept into a fundamental
notion in modern physics. Developments in statistical mechanics,
quantum information theory, and black-hole thermodynamics have revealed
deep connections between information, entropy, energy, and spacetime.
In particular, the recognition that information processing is subject
to thermodynamic constraints has established information erasure as an
important interface between information theory and physics
\cite{shannon1948,bennett1982}.

A fundamental result in this context is Landauer's principle, proposed
by Landauer in 1961 \cite{landauer1961irreversibility}. It states that
the logically irreversible erasure of one bit of information requires a
minimum heat dissipation determined by the temperature of the
surrounding environment. This principle provides a direct
thermodynamic connection between information and energy and has been
studied in classical and quantum computation, nonequilibrium
thermodynamics, and quantum information processing
\cite{bennett2003,parrondo2015,Herr}.

The interplay between information, thermodynamics, and gravity is
particularly evident in horizon physics. The discovery of black-hole
entropy by Bekenstein \cite{bekenstein1973} and Hawking radiation
\cite{hawking1975} established the thermodynamic nature of gravitational
horizons and motivated its extension to cosmological spacetimes, where
horizons can be associated with entropy and effective temperatures
determined by spacetime geometry
\cite{gibbons1977,cai2005,akbar2006}.

The Friedmann--Robertson--Walker (FRW) spacetime provides a natural
setting for studying this interplay. Its apparent horizon admits a
thermodynamic description closely related to gravitational dynamics,
and the Friedmann equations can be expressed in terms of horizon
entropy, temperature, and energy
\cite{cai2005,cai2009,Trivedi}. This raises the question of how
Landauer's principle should be interpreted when the apparent-horizon
temperature characterizes the effective cosmological environment.
Related connections between information, gravity, and cosmology have
also been explored through entropic gravity, holographic ideas, and
information-theoretic descriptions of cosmological horizons
\cite{verlinde2011,padmanabhan2010,padmanabhan2012}.

Recent studies have directly examined Landauer erasure in gravitational
and cosmological settings. Trivedi~\cite{Trivedi} found a factor-of-two
difference between the change in Misner--Sharp energy at the
cosmological apparent horizon and the standard Landauer cost, while
other recent works have investigated Landauer's principle during cosmic
evolution and in generalized gravitational thermodynamics
\cite{Odintsov2025,Elizalde2025,Ambrosio2025,Abreu2025}. These results
indicate that the relation between information erasure, horizon
thermodynamics, and gravitational energy depends on the thermodynamic
framework and on the physical system regarded as the information-bearing
system.

In this work, we adopt the standard apparent-horizon temperature as an effective cosmological temperature and investigate the Landauer heat required to erase one bit of information in spatially flat FRW cosmologies. We examine its dependence on the Hubble parameter, cosmic energy density, and horizon entropy, and compare it with the Misner--Sharp energy. We stress that these quantities have distinct physical meanings: the former is the minimum dissipative heat associated with irreversible erasure, whereas the latter is a quasi-local gravitational energy.

We find that, for standard non-phantom cosmological expansion, the Landauer heat decreases while the entropy-equivalent information content of the apparent horizon increases. We further show that the Misner--Sharp energy and horizon entropy satisfy a differential thermodynamic relation, clarifying why the Landauer heat should not be directly identified with gravitational energy and highlighting the role of the apparent-horizon temperature in connecting information erasure with cosmological horizon thermodynamics.

\section{Landauer principle and information erasure}
\label{Landauer principle}

The Landauer principle establishes a fundamental connection between
information theory and thermodynamics. It states that any logically
irreversible operation, such as the erasure of one bit of information,
must be accompanied by a minimum dissipation of heat into the surrounding
environment. For a thermal reservoir at temperature $T$, the minimum heat
required to erase one bit is given by

\begin{equation}
Q_{\rm L,min}=k_B T\ln 2.
\label{landauer}
\end{equation}
This result follows directly from the second law of thermodynamics.
Consider a memory device initially prepared in a maximally uncertain state
with two equiprobable logical configurations. The corresponding Shannon
entropy is

\begin{equation}
S_i=-k_B\sum_i p_i\ln p_i=k_B\ln2.
\end{equation}
After a reset operation, the memory is brought into a definite logical
state and its entropy vanishes, $S_f=0.$ Therefore, the entropy reduction associated with the erasure process is

\begin{equation}
\Delta S_{\rm info}=S_f-S_i=-k_B\ln2.
\end{equation}
According to the second law, the total entropy change of the memory plus
environment satisfies

\begin{equation}
\Delta S_{\rm tot}=\Delta S_{\rm info}+\Delta S_{\rm env}\geq 0.
\end{equation}
Consequently, the entropy increase of the environment must obey

\begin{equation}
\Delta S_{\rm env}\geq k_B\ln2.
\end{equation}
For a thermal reservoir at temperature $T$, the environmental entropy
variation is related to the heat dissipated into the reservoir through $\Delta S_{\rm env}=\frac{Q}{T}$, which immediately yields the Landauer bound

\begin{equation}
Q\geq k_B T\ln2,
\qquad
Q_{\rm L,min}=k_B T\ln2.
\end{equation}
It is important to distinguish the Landauer heat in Eq.~\eqref{landauer}
from the change in the energy associated with a gravitational horizon.
The former refers to the minimum heat that must be dissipated into the
environment of the memory during a logically irreversible reset. The
latter is a property of the gravitational spacetime and, in particular,
may be characterized by the variation of a quasi-local energy associated
with the horizon. These two quantities need not coincide and will be
carefully distinguished below.

The central question addressed in this paper is therefore how the
Landauer bound should be interpreted in an expanding cosmological
background, where the relevant temperature may acquire a geometric
origin associated with spacetime horizons. In particular, we investigate
the relation between the Landauer erasure cost and the energy variation
associated with the apparent horizon of an FRW universe.

From now on, we employ natural units $\hbar=c=k_B=1$ throughout this work.

\section{Effective horizon temperature in FRW cosmology}
\label{Effective horizon}

We now turn to the cosmological setting and investigate how the
Landauer bound may be applied in an expanding universe. The key
ingredient in the Landauer principle is the temperature of the
environment with which the information-bearing system exchanges
energy. In a FRW spacetime, however, the notion of temperature is
not unique. In addition to the temperature of ordinary matter or
radiation, cosmological horizons can possess thermodynamic
properties and can be assigned an effective temperature determined
by the spacetime geometry.

The line element of a homogeneous and isotropic FRW universe is
\begin{equation}
ds^{2}=-dt^{2}+a^{2}(t)\left(\frac{dr^{2}}{1-kr^{2}}+r^{2}d\Omega^{2}\right),
\end{equation}
where $a(t)$ is the scale factor and $k=0,\pm1$ denotes the spatial
curvature. Introducing the areal radius $\tilde r=a(t)r$, the
apparent horizon is defined by
$h^{ab}\partial_a\tilde r\,\partial_b\tilde r=0$,
where $h_{ab}$ is the metric on the $(t,r)$ submanifold. This
condition gives
\begin{equation}
R_A=\frac{1}{\sqrt{H^{2}+k/a^{2}}},
\label{RA}
\end{equation}
where $H=\frac{\dot a}{a}$ is the Hubble parameter.

The apparent horizon of an FRW universe admits a thermodynamic
description analogous in several respects to that of gravitational
horizons. In particular, an effective temperature associated with
the apparent horizon can be introduced through its surface gravity
\cite{cai2007, hayward1998}. In the quasi-static description commonly
used in FRW horizon thermodynamics, this temperature takes the form \cite{Rong}
\begin{equation}
T_A=\frac{1}{2\pi R_A}.
\label{TA}
\end{equation}
For a de Sitter universe, where $H$ is constant and $k=0$, one has
$R_A=H^{-1}$, and Eq.~\eqref{TA} reduces to the standard
Gibbons--Hawking temperature,
\begin{equation}
T_{GH}=\frac{H}{2\pi}.
\end{equation}
Equation~\eqref{TA} will be used here as an effective horizon
temperature. This choice should be distinguished from the temperature
of a material thermal bath and from the exact dynamical surface-gravity
temperature that can be defined for a time-dependent apparent horizon.
Indeed, for a genuinely dynamical FRW apparent horizon, the Hayward--Kodama surface gravity is instead
given by \cite{cai2009,hayward1998}

\begin{equation}
\kappa_{\rm HK}=-\frac{1}{R_A}\left(1-\frac{\dot R_A}{2H R_A}\right),
\end{equation}
which leads to the dynamical horizon temperature

\begin{equation}
T_{\rm HK}=\frac{|\kappa_{\rm HK}|}{2\pi}=\frac{1}{2\pi R_A}\left|1-\frac{\dot R_A}{2H R_A}\right|=\mathcal{F}T_A, 
\label{THK}
\end{equation}
in which we have introduced the dimensionless factor $\mathcal{F}\equiv\left|1-\frac{\dot R_A}{2H R_A}\right|$.
Thus, the quasi-static prescription adopted in the present work
corresponds to the limit
$\dot R_A/(2H R_A)\rightarrow0$, for which
$\mathcal{F}\rightarrow1$ and $T_{\rm HK}\rightarrow T_A$.
The distinction is important because the dynamical correction also
modifies the relations between the horizon temperature, entropy, and
Misner--Sharp energy, as discussed below. The main analysis of this
work uses the quasi-static temperature in Eq.~(\ref{TA}) as the
effective environmental temperature entering the Landauer bound.

Using Eq.~\eqref{TA}, the minimum heat dissipated in erasing one bit becomes
\begin{equation}
Q_{\rm L,min}=\frac{\ln 2}{2\pi}\sqrt{H^{2}+\frac{k}{a^{2}}}.
\end{equation}
This expression represents the Landauer heat associated with an
information-bearing system coupled to an environment characterized
by the effective temperature $T_A$. Importantly, it does not by
itself imply that this heat is equal to the change in the
gravitational energy associated with the apparent horizon. The latter
is a separate quantity that will be determined below from the
Misner--Sharp energy.

The apparent-horizon temperature is motivated by the geometrical and thermodynamical properties of FRW apparent horizons. Unlike a matter or radiation temperature, it is defined directly from the spacetime geometry and reduces to the Gibbons--Hawking temperature in the de Sitter limit. We therefore use $T_A$ as an effective temperature characterizing the cosmological environment, without identifying the horizon itself with the thermal reservoir in the operational formulation of Landauer's principle. The resulting Landauer heat is then compared with the quasi-local gravitational energy associated with the horizon.

\section{Landauer bound in FRW cosmology}
\label{Landauer bound}

Having identified the apparent-horizon temperature as an effective
temperature entering the Landauer principle, we now investigate the
corresponding erasure cost in spatially flat FRW cosmologies. For
$k=0$, the apparent-horizon radius and temperature reduce to $R_A=\frac{1}{H}$ and $T_A=\frac{H}{2\pi}$ respectively. 
Substitution into the Landauer bound gives

\begin{equation}
Q_{\rm L,min}=\frac{\ln 2}{2\pi}H \Rightarrow \frac{Q_{\rm L,min}}{H}=\frac{\ln 2}{2\pi}.
\label{general_landauer}
\end{equation}
Thus, within the effective-temperature prescription adopted here, the
minimum Landauer heat associated with erasing one bit is proportional
to the Hubble parameter. The dimensionless ratio in
Eq.~\eqref{general_landauer} is independent of the particular
cosmological matter content. This result reflects the geometrical
origin of the effective temperature used in the present framework.
The proportionality to $H$ also implies that the Landauer cost evolves
with the cosmological background. Consequently, different expansion
histories lead to different time dependences of the minimum erasure
heat, even though the ratio $Q_{\rm L,min}/H$ remains constant.

From the perspective of the relation with energy density, we consider a spatially flat FRW universe. The Friedmann equation is

\begin{equation}
H^2=\frac{8\pi G}{3}\rho.
\label{ Friedmann}
\end{equation}
Using this relation in Eq.~\eqref{general_landauer}, we obtain

\begin{equation}
Q_{\rm L,min}=\ln 2\sqrt{\frac{2G\rho}{3\pi}}\Rightarrow Q_{\rm L,min}\propto\sqrt{\rho}.
\label{rho_relation}
\end{equation}
Hence, within the present effective-temperature description, a larger
cosmic energy density corresponds to a larger minimum Landauer heat.
This dependence follows from the relation between the Hubble parameter
and the energy density and does not require a specific equation of
state for the cosmic fluid.

Now, assume that the mentioned universe is filled with a perfect fluid satisfying $p=w\rho$, where $w$ is constant. For $w\neq -1$, the scale factor evolves as

\begin{equation}
a(t)\propto
t^{\frac{2}{3(1+w)}},
\end{equation}
and the corresponding Hubble parameter is

\begin{equation}
H=
\frac{2}{3(1+w)t}.
\end{equation}
Substituting this expression into Eq.~\eqref{general_landauer} gives

\begin{equation}
Q_{\rm L,min}=\frac{\ln 2}{3\pi(1+w)}\frac{1}{t}.
\label{general_w}
\end{equation}
For the standard cases with $w>-1$, the minimum Landauer heat
therefore decreases as $t^{-1}$. Using the time dependence of the
scale factor, Eq.~\eqref{general_w} may alternatively be written as

\begin{equation}
Q_{\rm L,min}\propto a^{-\frac{3}{2}(1+w)}.
\label{scale_factor_relation}
\end{equation}
Thus, for the usual non-phantom cosmological fluids, the energetic
cost associated with the effective horizon temperature decreases as
the universe expands. The precise scaling is determined by the
equation of state of the dominant cosmic component.

For a de Sitter universe, $w=-1$, the scale factor evolves
exponentially,

\begin{equation}
a(t)=e^{H t},
\end{equation}
with constant Hubble parameter $H$. In this case, $T_A=\frac{H}{2\pi}$, and the Landauer bound becomes

\begin{equation}
Q_{\rm L,min}^{(\mathrm{dS})}=\frac{H\ln 2}{2\pi}.
\end{equation}
The minimum Landauer heat is therefore constant, reflecting the
stationary horizon and constant horizon temperature of de Sitter
spacetime.

Finally, in viewpoint of the horizon entropy, note that the apparent horizon possesses an entropy described, in the standard
geometrical formulation, by the Bekenstein--Hawking area law,

\begin{equation}
S_A=\frac{A}{4G},
\end{equation}
where $A=4\pi R_A^2$. For a spatially flat universe, $R_A=H^{-1}$, and hence

\begin{equation}
S_A=\frac{\pi}{G H^2}.
\label{horizon_entropy}
\end{equation}
Solving Eq.~\eqref{horizon_entropy} for the Hubble parameter and
substituting the result into Eq.~\eqref{general_landauer}, we obtain

\begin{equation}
Q_{\rm L,min}
=
\frac{\ln 2}{2}
\sqrt{\frac{1}{\pi G S_A}} \Rightarrow Q_{\rm L,min}\propto S_A^{-1/2}.
\label{entropy_relation}
\end{equation}
The Landauer heat associated with erasing one bit therefore decreases
as the entropy of the apparent horizon increases. Combining
Eqs.~\eqref{general_landauer}, \eqref{rho_relation}, and
\eqref{entropy_relation}, we may summarize the corresponding scaling
relations as

\begin{equation}
Q_{\rm L,min}\propto H\propto\sqrt{\rho}\propto S_A^{-1/2}.
\label{scaling_relations}
\end{equation}
These relations summarize the dependence of the single-bit Landauer
cost on the expansion rate, cosmic energy density, and horizon entropy.
They characterize the heat cost obtained from the Landauer prescription
and do not identify it with the variation of the gravitational energy
associated with the apparent horizon, which is considered next.

\section{Horizon information and gravitational energy}
\label{Horizon information}

The cosmological dependence of the Landauer heat obtained above
naturally distinguishes the early and late stages of cosmic evolution.
For the standard non-phantom cosmological histories considered here,
the decrease of $H$ implies a decreasing single-bit Landauer cost, while
the horizon entropy grows. We now characterize the corresponding
entropy-equivalent information content and compare the Landauer scale
with the Misner--Sharp energy.

The apparent horizon carries an entropy (\ref{horizon_entropy}), which allows one to associate an entropy-equivalent information content with the horizon. The corresponding effective number of horizon bits associated with the entropy $S_A$ is 

\begin{equation}
N_A=\frac{S_A}{\ln2}.
\label{NA}
\end{equation}
Here, $N_A$ is an information-
theoretic interpretation of the Bekenstein--Hawking entropy rather than
a demonstrated operational storage capacity of the cosmological
horizon. No microscopic encoding mechanism, quantum state-counting
prescription, or physical memory model for the horizon is assumed in
this definition. For the spatially flat FRW universe, this gives

\begin{equation}
N_A=\frac{\pi}{G H^2\ln2}.
\end{equation}
Thus, while the minimum Landauer heat per bit decreases as the universe
expands, the effective number of horizon bits associated with the apparent horizon
increases as its area grows. This provides a useful distinction between
the microscopic cost of erasing a single bit and the entropy-equivalent information content of the horizon.

It is useful at this point to examine the gravitational energy
associated with the same horizon. The Misner--Sharp energy enclosed
within a spherically symmetric region is

\begin{equation}
E_{\rm MS}=\rho V,
\end{equation}
where, for a spatially flat FRW universe, $V=\frac{4\pi}{3H^3}$ is the horizon volume.
Using the Friedmann equation (\ref{ Friedmann}), one obtains

\begin{equation}
E_{\rm MS}
=
\frac{1}{2GH}.
\label{EMS}
\end{equation}
The same result follows directly from the geometrical expression for
the Misner--Sharp energy evaluated at the apparent horizon,

\begin{equation}
E_{\rm MS}=\frac{R_A}{2G}.
\end{equation}
For $R_A=H^{-1}$, this again gives Eq.~\eqref{EMS}.

An interesting relation follows immediately from the horizon entropy
and temperature,

\begin{equation}
S_A T_A
=
\frac{\pi}{GH^2}\frac{H}{2\pi}
=
\frac{1}{2GH}.
\end{equation}
Consequently,

\begin{equation}
E_{\rm MS}=S_A T_A.
\label{EST}
\end{equation}
Equation~\eqref{EST} is a thermodynamic-geometrical identity for the
spatially flat FRW apparent horizon. It should not, however, be
interpreted as an independent derivation of the Landauer erasure
energy. Indeed, multiplying the number of horizon bits by the
single-bit Landauer bound gives

\begin{equation}
N_A Q_{\rm L,min}=\frac{S_A}{\ln2}\,T_A\ln2=S_A T_A=E_{\rm MS}.
\label{formal_scale}
\end{equation}
This equality follows algebraically from the definitions of
$N_A$, $Q_{\rm L,min}$, $S_A$, and $T_A$. It therefore does not imply
that the dynamical erasure of all information carried by the
cosmological horizon necessarily requires an energy equal to the
Misner--Sharp energy. In particular, the Landauer heat is the minimum
heat dissipated into an environment during an irreversible reset,
whereas $E_{\rm MS}$ is a quasi-local gravitational energy associated
with the spacetime geometry.

The distinction becomes particularly important when considering a
change in the horizon entropy. From (\ref{horizon_entropy}) and (\ref{EMS}), the Misner--Sharp energy scales with the horizon entropy as

\begin{equation}
E_{\rm MS}=\frac{1}{2\sqrt{\pi G}}\sqrt{S_A}.
\label{MS1}
\end{equation}
Consequently,

\begin{equation}
\frac{dE_{\rm MS}}{dS_A}=\frac{1}{4\sqrt{\pi G S_A}}.
\end{equation}
Using

\begin{equation}
T_A=\frac{1}{2\sqrt{\pi G S_A}},
\label{TAA}
\end{equation}
we obtain the important differential relation

\begin{equation}
dE_{\rm MS}=\frac{1}{2}T_A\,dS_A.
\label{dEMS}
\end{equation}
This equation is valid for infinitesimal changes in the horizon entropy. Since
$E_{\rm MS}\propto\sqrt{S_A}$, however, a one-bit change
$\Delta S_A=\ln2$ is finite and should be treated using the exact
finite difference. From (\ref{MS1}), the change associated with an increase of one bit is exactly

\begin{equation}
\Delta E_{\rm MS}=\frac{1}{2\sqrt{\pi G}}\left[\sqrt{S_A+\ln2}-\sqrt{S_A}\right].
\label{exact_delta_EMS}
\end{equation}
Equivalently, with the help of (\ref{TAA}), this can be written as

\begin{equation}
\Delta E_{\rm MS}=\frac{1}{2}T_A\ln2\left[\frac{2}{1+\sqrt{1+\frac{\ln2}{S_A}}}\right].
\label{exact_delta_EMS_T}
\end{equation}
Therefore, the exact finite one-bit change is not identically equal to
the differential result. In the large-entropy limit
$S_A\gg\ln2$, the correction factor approaches unity and

\begin{equation}
\Delta E_{\rm MS}=\frac{1}{2}T_A\ln2\left[1-\frac{\ln2}{4S_A}+\mathcal{O}\left(\frac{(\ln2)^2}{S_A^2}\right)\right].
\end{equation}
Thus, for a cosmological horizon with $S_A\gg\ln2$, the corresponding magnitude of the Misner--Sharp energy variation reduces to

\begin{equation}
|\Delta E_{\rm MS}|\simeq\frac{1}{2}T_A\ln2.
\label{DeltaEMS}
\end{equation}
This result differs by a factor of two from the standard Landauer
heat,

\begin{equation}
Q_{\rm L,min}=T_A\ln2.
\label{QLagain}
\end{equation}
The two quantities therefore characterize different aspects of the
problem. The Landauer bound refers to the minimum heat dissipated by
an information-bearing system coupled to an environment at temperature
$T_A$, whereas Eq.~\eqref{DeltaEMS} describes the variation of the
quasi-local gravitational energy resulting from a change in the
horizon entropy. The factor of $1/2$ follows directly from the
different dependence of $E_{\rm MS}$ and $S_A$ on the Hubble parameter,
and should not be interpreted as a modification of the standard
Landauer bound.

This distinction provides a natural way to reconcile the two energy
scales. Rather than identifying the Landauer heat with the variation
of the Misner--Sharp energy, the present framework relates them through

\begin{equation}
Q_{\rm L,min}=2|\Delta E_{\rm MS}|
\end{equation}
for a one-bit change in the horizon entropy, within the linearized
description above.

As a final remark, note that the relation $E_{\rm MS}=S_A T_A$, is prescription
dependent, however, and is modified when the fully dynamical
Hayward--Kodama temperature is employed. Using
$T_{\rm HK}=\mathcal{F}T_A$, one instead obtains

\begin{equation}
S_A T_{\rm HK}=\mathcal{F}E_{\rm MS}\Rightarrow E_{\rm MS}=\frac{S_A T_{\rm HK}}{\mathcal{F}}.
\end{equation}
The differential relation is modified in the same way as

\begin{equation}
T_{\rm HK}dS_A=\mathcal{F}\frac{dR_A}{G},
\end{equation}
from which we obtain, for $\mathcal{F}\neq0$,

\begin{equation}
dE_{\rm MS}=\frac{1}{2\mathcal{F}}T_{\rm HK}\,dS_A.
\label{dEMS_dynamic}
\end{equation}
Consequently, for a one-bit change in the horizon entropy,

\begin{equation}
|\Delta E_{\rm MS}|=\frac{1}{2\mathcal{F}}T_{\rm HK}\ln2,
\end{equation}
to first order. The factor-of-two relation obtained in the main
analysis is therefore recovered in the quasi-static limit
$\mathcal{F}\rightarrow1$. Away from this limit, the relation between
the Misner--Sharp energy variation and the Landauer scale acquires the
additional dynamical factor $1/\mathcal{F}$.

\subsection{Cosmological examples}

To illustrate the different evolution of the single-bit Landauer heat
and the horizon information content, we consider three representative
cosmological models: a radiation-dominated universe, a
matter-dominated universe, and a de Sitter universe.

For the radiation-dominated case $(w=1/3)$, the relevant quantities
are

\begin{equation}
H=\frac{1}{2t},
\qquad
Q_{\rm L,min}=\frac{\ln2}{4\pi t},
\qquad
N_A=\frac{4\pi t^2}{G\ln2},
\qquad
E_{\rm MS}=\frac{t}{G}.
\end{equation}
For a matter-dominated universe $(w=0)$, one obtains

\begin{equation}
H=\frac{2}{3t},
\qquad
Q_{\rm L,min}=\frac{\ln2}{3\pi t},
\qquad
N_A=\frac{9\pi t^2}{4G\ln2},
\qquad
E_{\rm MS}=\frac{3t}{4G}.
\end{equation}
Finally, for de Sitter spacetime,

\begin{equation}
H=H_0,
\qquad
Q_{\rm L,min}=\frac{H_0\ln2}{2\pi},
\qquad
N_A=\frac{\pi}{G H_0^2\ln2},
\qquad
E_{\rm MS}=\frac{1}{2GH_0}.
\end{equation}
In the radiation- and matter-dominated cases, the minimum Landauer
heat per bit decreases with cosmic time, whereas the number of bits
associated with the apparent horizon grows quadratically with time.
The Misner--Sharp energy, on the other hand, grows linearly with time.
For de Sitter spacetime, all three quantities remain constant.

\begin{figure}[t]
\centering
\begin{subfigure}[b]{0.4\textwidth}

\includegraphics[width=\textwidth]{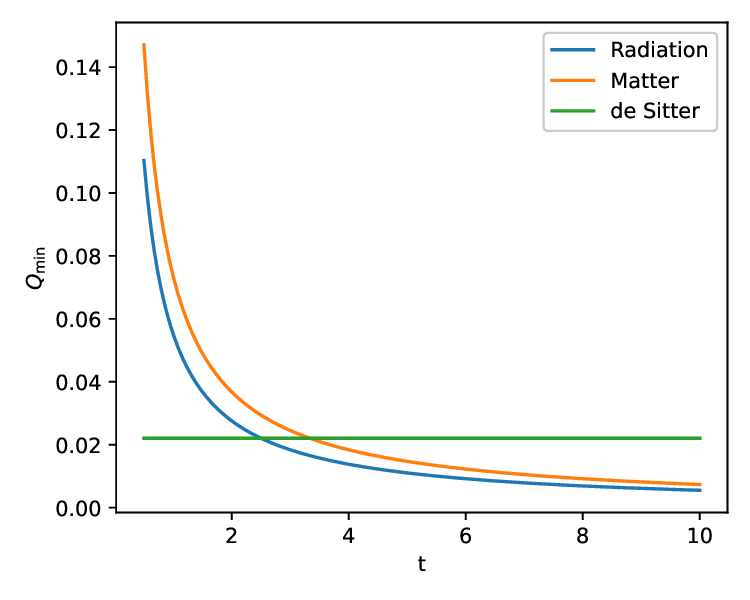}
\caption{}
\label{fig1(a)}
\end{subfigure}
\hspace{5mm}
\begin{subfigure}[b]{0.4\textwidth}

\includegraphics[width=\textwidth]{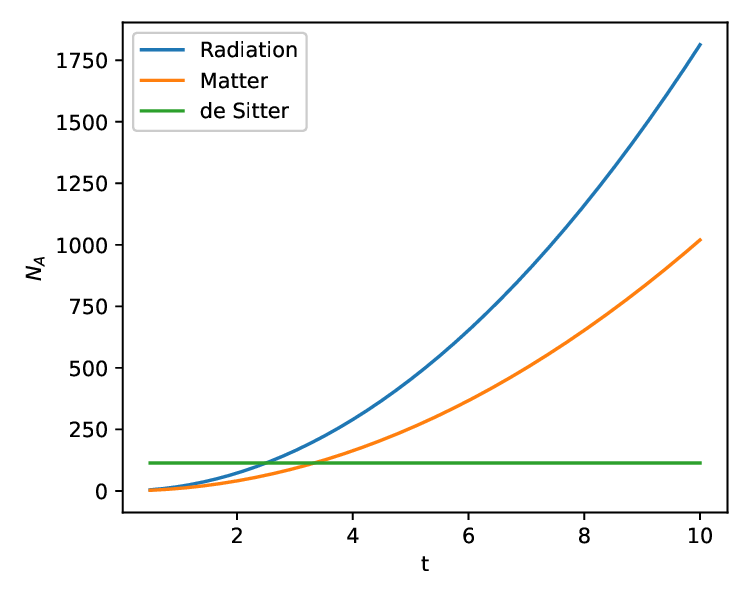}
\caption{}
\label{fig1(b)}
\end{subfigure}
\hspace{5mm}
\begin{subfigure}[b]{0.4\textwidth}

\includegraphics[width=\textwidth]{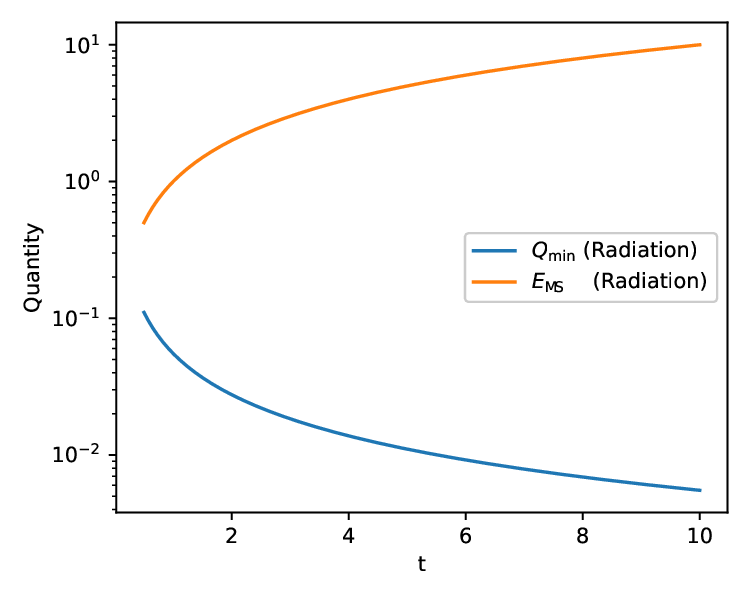}
\caption{}
\label{fig1(c)}
\end{subfigure}
\caption{(a) Evolution of the minimum Landauer heat required to
erase one bit of information, $Q_{\rm L,min}$, for radiation-dominated,
matter-dominated and de Sitter cosmologies. (b) Evolution of the
horizon information content, $N_A=S_A/\ln2$, for the same cosmological
models. (c) Logarithmic comparison of the minimum Landauer heat per
bit, $Q_{\rm L,min}$, and the Misner--Sharp energy $E_{\rm MS}$ in a
radiation-dominated universe. The figures are plotted in normalized
units with $G=1$.}
\label{fig1}
\end{figure}

The numerical behavior of these quantities is summarized in Fig.~\ref{fig1}.
As Figs. \ref{fig1(a)} and \ref{fig1(b)} show, for radiation- and matter-dominated cosmologies, $Q_{\rm L,min}$
decreases while $N_A$ and $E_{\rm MS}$ increase with cosmic time. In
de Sitter spacetime, all three quantities remain constant. Figure \ref{fig1(c)}
also illustrates the different physical roles of the single-bit
Landauer heat and the total quasi-local gravitational energy.

In summary, the cosmological evolution exhibits a competition between
the decreasing cost of erasing an individual bit and the increasing
information capacity of the apparent horizon. At the same time, the
Misner--Sharp energy associated with the horizon evolves independently
according to the gravitational dynamics. The comparison between these
quantities leads to the differential relation in Eq.~\eqref{dEMS},
which provides the basis for distinguishing the Landauer heat from the
horizon energy variation.

\section{Conclusions}
\label{Conclusions}

We have investigated the thermodynamic cost of information erasure in
a spatially flat FRW universe by combining Landauer's principle with
the thermodynamics of the apparent horizon. Using the standard
quasi-static apparent-horizon temperature as an effective cosmological
temperature, we find $Q_{\rm L,min}\propto H\propto \sqrt{\rho}\propto S_A^{-1/2}$.

Thus, for standard non-phantom FLRW expansion, the single-bit
Landauer cost decreases while the entropy-equivalent information
content of the apparent horizon increases. De Sitter spacetime
provides the limiting case in which both quantities remain constant.

We have also compared the Landauer heat with the Misner--Sharp energy.
For the quasi-static temperature,
$E_{\rm MS}=S_A T_A$, but this equality does not identify the two
quantities physically: the Landauer bound is a minimum dissipative heat
for an irreversible erasure, whereas $E_{\rm MS}$ is a quasi-local
gravitational energy. Their differential relation is
$dE_{\rm MS}=\frac12 T_A dS_A$. Treating one bit as a finite entropy
change, $\Delta S_A=\ln2$, gives the exact finite-difference expression
derived in the text and reduces to
$|\Delta E_{\rm MS}|\simeq\frac12T_A\ln2$ in the
$S_A\gg\ln2$ limit. The resulting factor of two therefore reflects the
distinct physical meanings of the two energy scales rather than a
modification of Landauer's principle.

We have further shown how these relations are modified when the fully
dynamical Hayward--Kodama temperature is used. The present analysis
adopts the quasi-static prescription as the effective environmental
temperature, while the dynamical treatment provides the corresponding
correction and clarifies the domain of the main results.

The framework developed here can be extended to non-flat FRW
geometries, dynamical dark-energy models, and quantum-corrected
cosmological horizons. Such extensions may also allow genuinely
quantum-information measures to be incorporated into the
thermodynamic description of information processing in curved
spacetime.

\end{document}